\documentclass[preprint2]{aastex}

\begin{document}
\newcommand{\refp}[1]{(\ref{#1})}

\title{The potential impact of groove modes on Type II planetary migration}
\author{Stefano Meschiari, Gregory Laughlin}
\affil{Dept. of Astronomy \& Astrophysics, University of California,
                Santa Cruz, CA 95064}

\shorttitle{Groove modes}
\shortauthors{Meschiari \& Laughlin}

\begin{abstract}
In this letter, we briefly describe the evolution of a variety of self-gravitating protoplanetary disk models that contain annular \emph{grooves} (e.g. gaps) in their surface density. These grooves are inspired by the density gaps that are presumed to open in response to the formation of a giant planet. Our work provides an extension of the previously studied \emph{groove modes} that are known in the context of stellar disks. The emergence of spiral gravitational instabilities (GI) is predicted via a generalized eigenvalue code that performs a linear analysis, and confirmed with hydrodynamical simulations. We find the presence of a groove drives a fast-growing two-armed mode in moderately massive disks, and extends the importance of self-gravitating instabilities down to lower disk masses than for which they would otherwise occur. We discuss the potential importance of this instability in the context of planet formation, e.g. the modification of the torques driving Type II migration.
\end{abstract}


\newcommand{\plotonespec}[1]{\centering{\epsscale{0.90}\plotone{#1}}}
\newcommand{\plotonespecb}[1]{\centering{\epsscale{0.55}\plotone{#1}}}
\newcommand{\plotonespecc}[1]{\centering{\epsscale{0.49}\plotone{#1}}}

\section{Introduction}
The theory of giant planet formation has evolved very rapidly over the past decade, and the now near-paradigmatic core accretion theory \citep[e.g.][]{Pollack1996} has been greatly refined and endowed with a variety of physical inputs \citep[see, e.g.][]{Alibert2004, Hubickyj2005}. It is now generally accepted that Jupiter-mass planets grow by initially accumulating 5-10 $M_\oplus$ cores, which then accrete large quantities of gas from the surrounding nebula. During the rapid gas accretion phase, the growing Jovian planet opens a gap in the disk and its orbit subsequently evolves via the process of Type II migration \citep{Lin1996, Papaloizou07}.  

A competing theory \citep[e.g.][]{Boss2000} holds that giant planets such as Jupiter fragmented directly from the nebula as a result of gravitational instabilities. This mechanism may account for some of the observed extrasolar planets, but it has a number of difficulties if invoked to account for the full population. In particular, it fails to explain the observationally well established planet-metallicity relation \citep{Santos03, FischerValenti2005}. Another criticism is that GI requires the Toomre $Q$ parameter, $Q = \kappa c_s / \pi G \sigma$, to be of order unity at some radius in the disk. This, however, generally requires rather massive disks, $M_D/M_{star} \gtrsim 0.1$, which are not often observed.

Global GIs depend not just on disk mass and sound speed, but also on the density profile of the disk. As shown by \citet{Toomre81}, a sharp density gradient (i.e. ``edges'') can drive a wave cycle that leads to the subsequent destruction of the edge. Analogously, \cite{SellwoodLin89} [SL89] described a family of GIs in stellar disks dubbed \emph{groove modes} which comprise fast-growing disturbances driven by narrowly defined structures in particle angle-action space. These features, in physical space, correspond to narrow density depressions; groove modes are thus akin to the edge mode in that they are driven by steep density gradients at corotation. SL89 further envisioned a feedback cycle which kicks in when a slowly-growing intrinsic mode (i.e. an inner edge mode) carves a groove by scattering particles in a narrow range at its Lindblad resonance and creating a fast-growing groove mode; the new groove gives rise to further groove modes once it travels to its own Lindblad resonances. 

This feedback cycle relies on the wave-particle interaction at the resonances, and thus cannot be immediately realized in a gaseous disk. The groove mode may have relevance, however, in the protoplanetary context: theoretical considerations \citep{LinPapaloizou79, GoldreichTremaine80, LinPapaloizou93}, numerical simulations \citep[e.g.][]{Takeuchi96, Bryden99} and tentative observational indications \citep{Setiawan08} point to the fact that massive planets can resonantly drive trailing waves that transport angular momentum and open a ``gap'' in the disk. The detailed process of opening and maintaining the gap depends on the balance between the angular momentum flux resulting from spiral waves driven by the planet and that due to the viscosity of the disk. When the planet is sufficiently massive and the viscosity is low enough, a stable groove can be naturally carved and mantained on the disk.

In this letter, we open an investigation into the possibility that the planet-induced density gap can drive a fast-growing, self-gravity induced groove mode in a disk, sidestepping the problems of carving a groove in the first place and providing a mechanism for maintaining the amplifier. We support our hypothesis with both a linear mode analysis \citep[see, e.g.][]{ARS1989} and a full hydrodynamical simulation \citep[following][hereafter LR96]{LaughlinRozyczka96} of a disk with an imposed groove in surface density, as inspired by previous numerical investigations for stellar disks \citep[e.g.][]{SellwoodKahn91}. We span a range of mass ratios to assess the relevance of this GI to the realistic disk masses observed.

\section{Procedure}
We employ a two-dimensional hydrodynamical grid code for following the evolution of a thin, self-gravitating disk. The continuity and Euler equations in polar coordinates are solved using a second-order van Leer type scheme, coupled with time stepping that is first-order accurate. The basic difference equations are given in \citet{StoneNorman1992}. The self-gravity of the disk is obtained by applying the Fourier convolution theorem to the potential dictated by the Poisson equation \citep{BinneyTremaine87}. The details of the hydrodynamical code are described in \citet{Laughlin97} and related papers.

We adopt the following parametrization for the surface density of the disk:
\begin{equation}
\sigma(r) = \sigma_0 e^{\left[-(r - R_0)^2/w\right]} \times
	\left(1 -\frac{A\Delta^2}{(r - R_P)^2 + \Delta^2}\right) ,
\label{eqn:surfacedensity}
\end{equation}
which represents a Gaussian profile multiplied by a
Lorentzian gap of depth $A$, semi-width $\Delta$ and central position
$R_P$; for $A = 0$, this profile is the ``reference disk'' considered by LR96. The disk model used throughout this letter has no pretense of actually representing a protoplanetary disk faithfully, but has the advantage of possessing a single intrinsic $m=2$ mode for $A = 0$ that is clearly identifiable both in semianalytic calculations and in the nonlinear simulations. The choice of a Lorentzian profile to represent the gap is arbitrary as well, and follows \citet{SellwoodKahn91}.  We take $w = 0.03$, $R_0 = 0.25$ and an inner edge of 0.05. A polytropic equation of state is assumed, $P = K \sigma^\gamma$ with $\gamma = 2$ (again, this is chosen largely for illustrative purposes). The characteristic width $\Delta$ is meant to represent a typical gap width; the chosen value of 0.07 can be derived from the approximate scaling derived by \citet[e.g.][]{Varniere04} with  $\mathcal{R} \sim 5\times10^5$ and $q = 2\times 10^{-3}$. 
The set of units used in the code takes the outer grid radius $R_D$ and the gravitational constant $G$ equal to unity and $M_* = 0.5$. 

To quantify the strength, pattern speed and growth rate 
of each spiral mode, we compute the Fourier decomposition of the surface density, defined as 
\begin{equation}
a_m = \frac{1}{2\pi}\int_0^{2\pi} \sigma(r, \Phi) e^{-i m \Phi}\mathrm{d}\Phi = c_m \times a_0, 
\end{equation}
for a mode number $m$. A global measure
of the growth of a particular mode is given by integrating the $m$-th Fourier amplitude $a_m$ over the radial range 
and normalizing to the azimuthal average of the surface density:
\begin{equation}
C_m = \left|\frac{\int_{R_i}^{R_D} a_m(r) dr}{\int_{R_i}^{R_D} a_0(r) dr}\right|, \ \gamma_m = \frac{\mathrm{d}}{\mathrm{d}t} \log C_m.
\end{equation}
The phase of a disturbance is recovered as $\Phi_m(r) = \tan^{-1} \left[\mathrm{Im}(-a_m)/\mathrm{Re}(a_m)\right]$; the local pattern speed is then given by $\Omega_P = (1/m) \dot{\Phi}_m$. 

\begin{figure}
\plotonespec{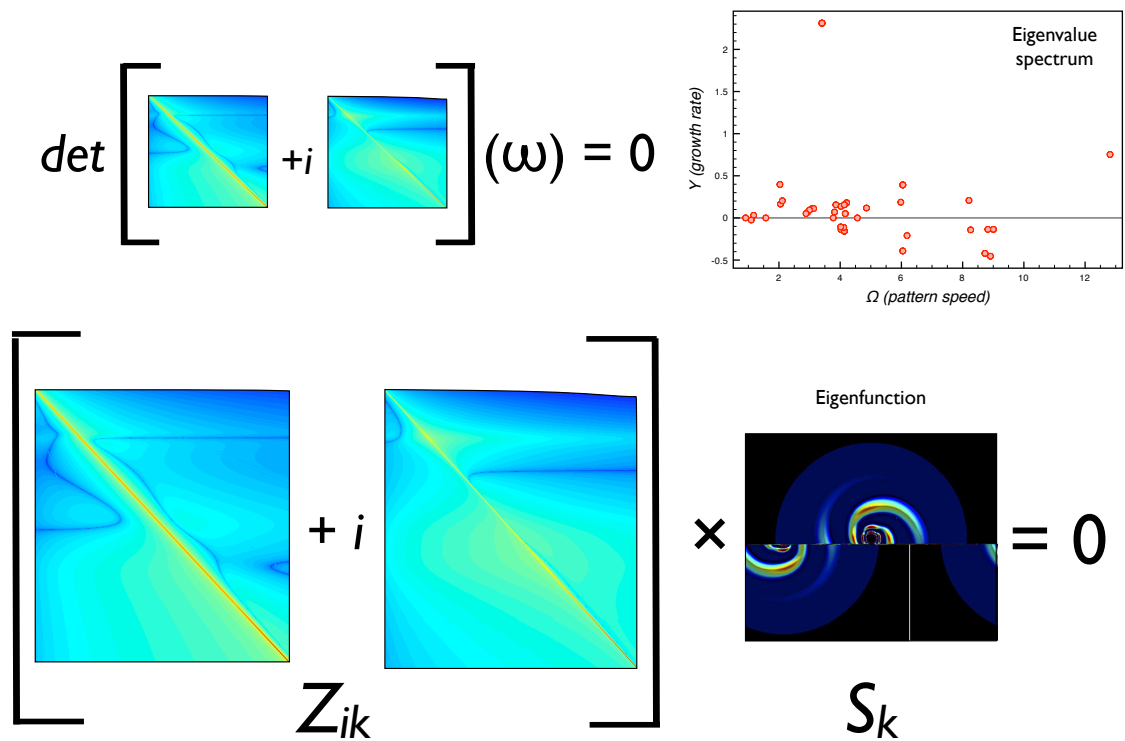}
\caption{Graphical depiction of the generalized eigenvalue problem. The square matrices represent the real and complex parts of the governing matrix. Solving the eigenvalue problem predicts the pattern speed, growth rate and radial dependence of the unstable mode.}
\label{fig:linear}
\end{figure}

The growth rate and pattern speed that emerge from the hydrodynamical simulation are checked against
a linear numerical analysis code we developed, as described e. g. in \citet{Laughlin97}, which solves a matrix equation akin
to a generalized eigenvalue problem; the solution is valid in the linear regime and yields 
a complex eigenvalue, which indicates the pattern speed $\Omega_P$ and growth rate $\gamma$, and a complex eigenvector, which describes the radial variation and phase of the mode. A graphical depiction of the procedure is shown in Figure \refp{fig:linear}. Comparison with the full nonlinear simulations enables us to check the consistency and accuracy of the two independent approaches.

\section{Computer simulations}

\begin{figure}
\plotonespec{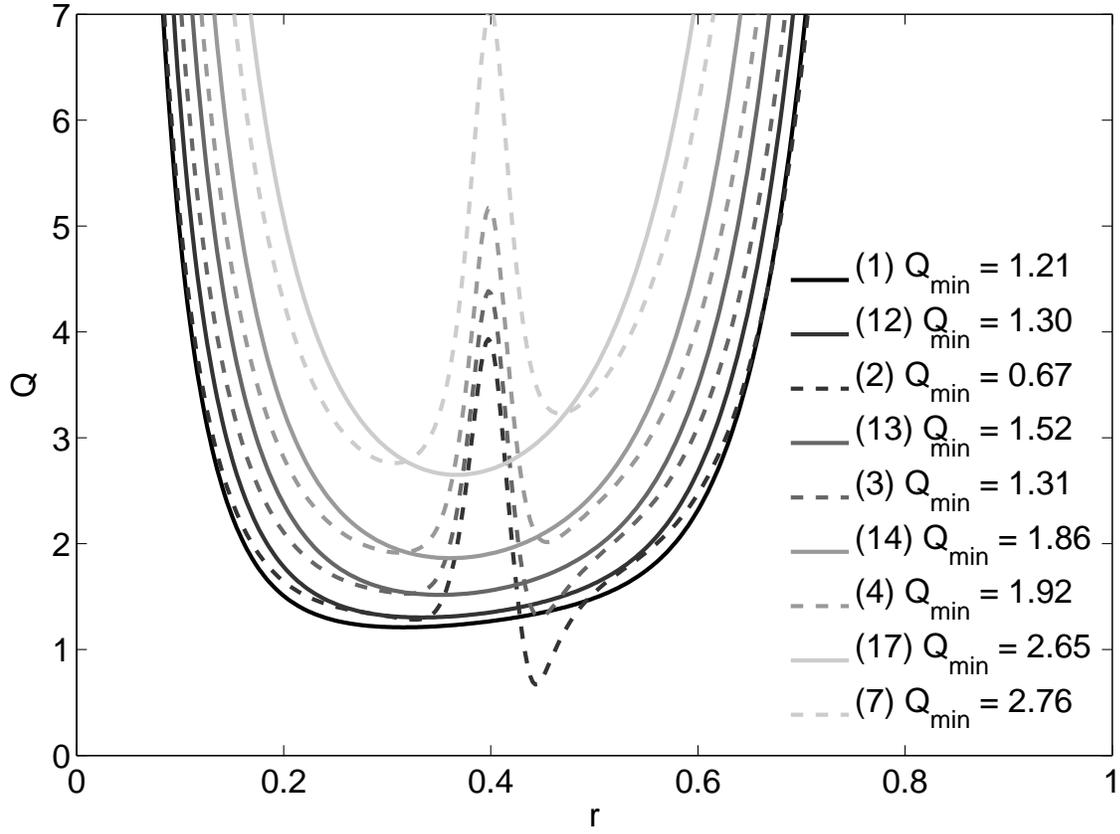}
\caption{Toomre $Q = \kappa c_s / \pi G \sigma$ profiles for the disk models considered (dashed lines: models with a groove, solid: models without a groove). Like colors correspond to equal $q_D$.}\label{fig:qs}
\end{figure}

Table \refp{tab:models} lists the disk models considered in this letter. We have first set up a ``base'' ($A = 0$) disk in rotational equilibrium with surface density given by Equation \ref{eqn:surfacedensity}, with
$q_D = M_D / M_{star} = 1$ and $Q_{min}=1.21$ (Model 1); this sets the normalization constant $\sigma_0$ and the polytropic constant $K$. The resulting Toomre $Q$ profiles are shown in Figure \refp{fig:qs}. The equilibrium was disrupted with a random 
density perturbation of order $10^{-3} \sigma(r)$. The grid covers the polar coordinates ($r, \varphi$) with 256 logarithmically spaced zones in radius and 256 equally spaced azimuthal zones. Each model is evolved for at least 100 time units, although the reflective boundaries muddle the nonlinear evolution once the wave reaches the outer part of the disk.

\begin{figure}
\plotonespec{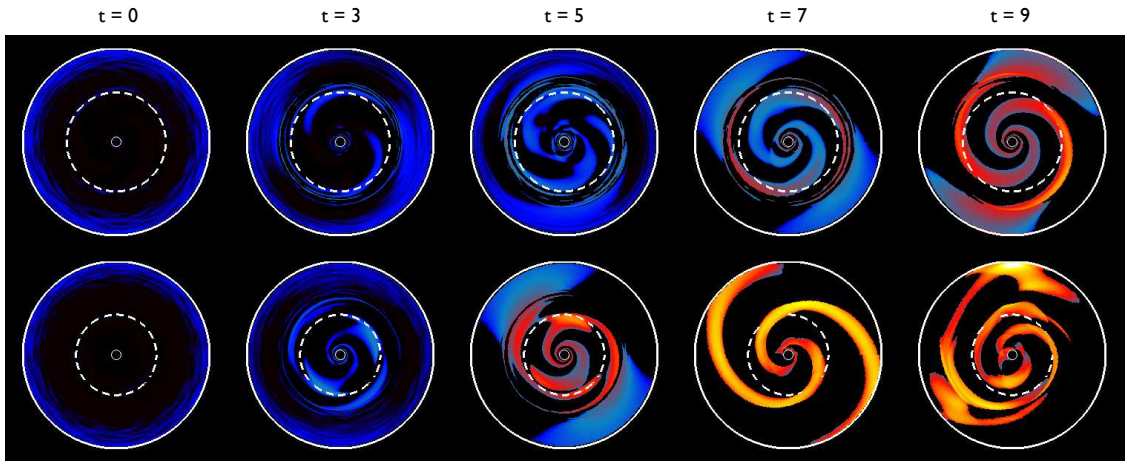}
\caption{Surface overdensity $\Sigma$ evolution for Models 1 and 2.}\label{fig:sd_groove}
\end{figure}

\begin{figure}
\plotonespec{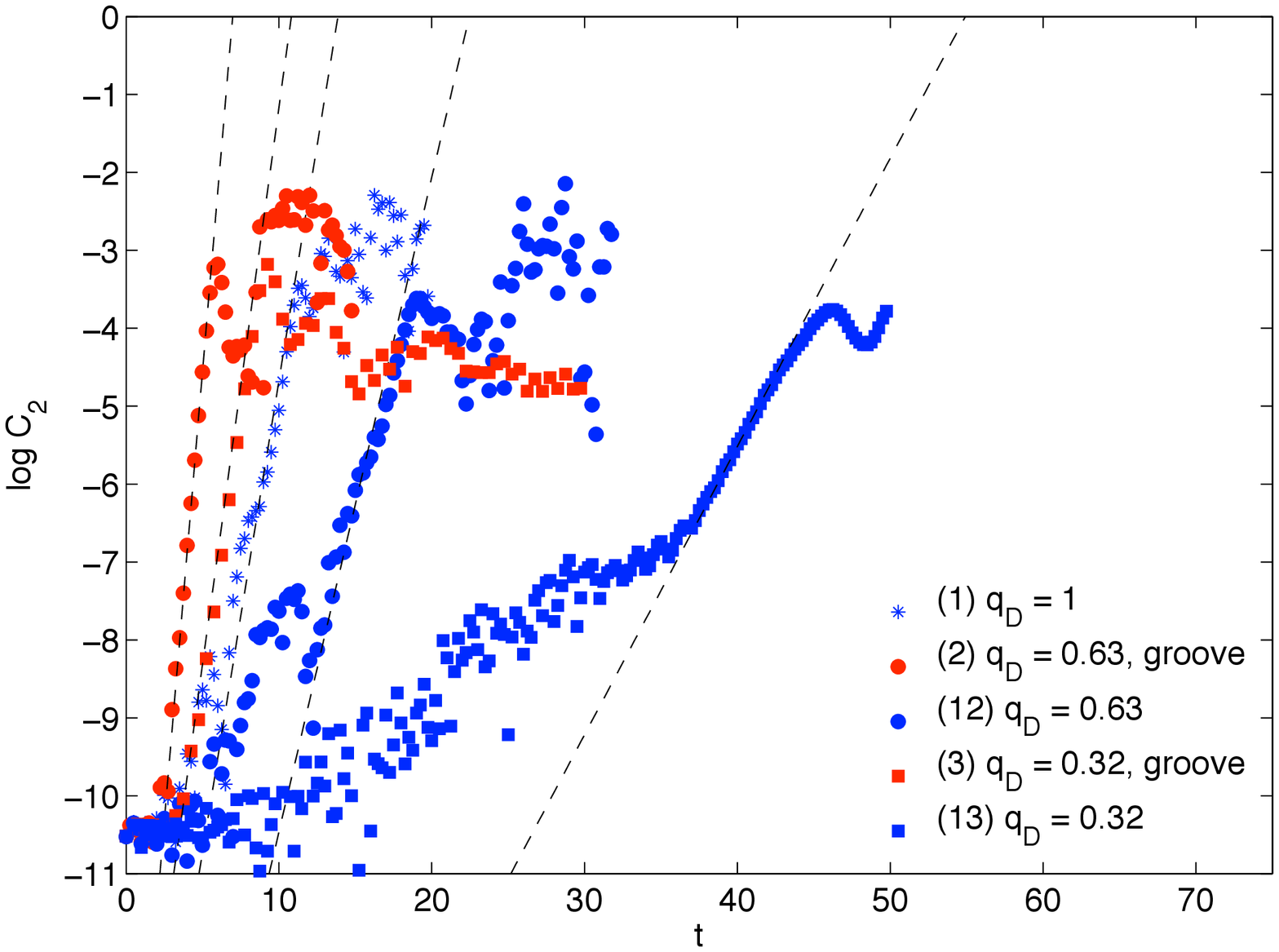}
\plotonespec{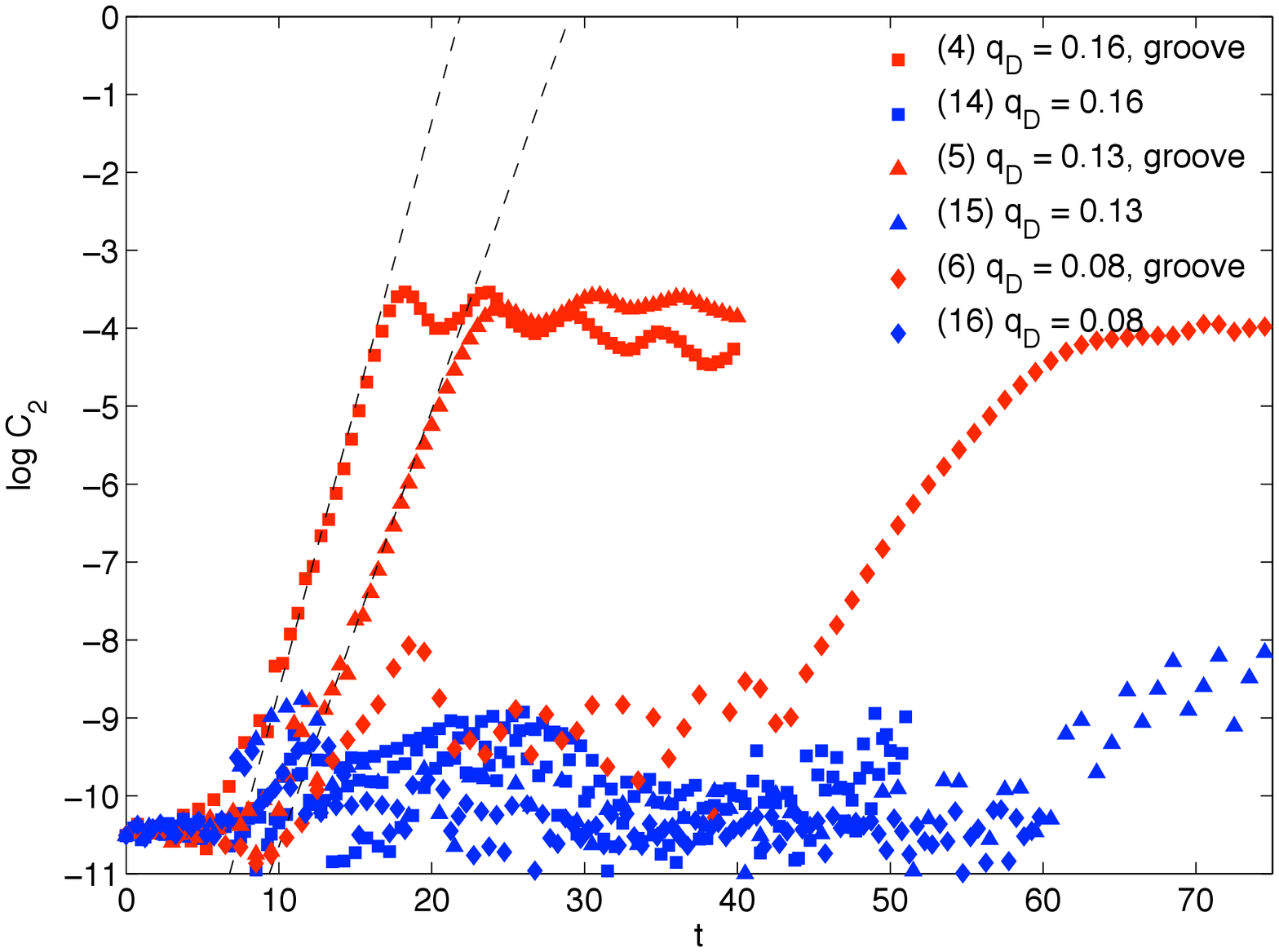}
\caption{Top panel: normalized amplitudes of the $m=2$ mode for disk models 1-3, 12, 13. Bottom panel: normalized amplitudes of the $m=2$ mode for disk models 4-6, 14-16. The slope of the dashed lines shows the predicted growth rate of a mode as resolved by our linear mode code.}\label{fig:evol}
\end{figure}

Figure \refp{fig:sd_groove} shows the evolution of the surface density normalized to the azimuthal average $\Sigma(r) = \log\left(\sigma(r) / \bar{\sigma}(r)\right)$. In accordance with the previous investigations and with our linear code, the base disk is unstable to a single $m=2$ ``grand-design'' spiral mode. The GI grows in the linear regime for the first few dynamical times, and within about 10 dynamical times it visibly perturbs the outer edges of the disk. After a few more dynamical times, the spiral pattern has reflected off the boundaries and has propagated back into the densest regions of the disk. The normalized amplitude $C_2$ (Figure \ref{fig:evol}) shows that the dominant two-armed mode grows at an exponential rate (linear regime) until stabilizing around a constant amplitude (mode saturation). For our purposes, the linear growth rate is the primary quantity of interest. 

A second set of disk parameters (Model 2) including a density \emph{groove} was set up, with $A = 0.90$ (a 90\% dip), $\Delta = 0.07$ and $R_P = 0.4$. Maintaining the same normalizations for density and pressure as above, the disk has the same density profile sufficiently far from the groove, but a smaller disk-to-star mass ratio, $q_D = 0.63$.  By visual inspection of Figure \refp{fig:sd_groove} and the slope of the normalized Fourier amplitude in Figure \refp{fig:evol}, it is clear that the presence of the groove drives a far more violently growing instability. The dashed linear slope is derived from the mode analysis for the same disk parameters, and agrees  well with the hydrodynamical code; the small discrepancies between the two methods can be traced to the effective softening given by the FFT-based solution  scheme for the Poisson equation. We then varied the density normalization to yield $q_D$ = 0.32, 0.16, 0.13, 0.08, and 0.06 (Models 3-11).

Measuring the pattern speed is more complex, since the spiral rotates only approximately in a rigid fashion. We thus calculate a ``global'' measure $\bar{\Omega}_P$ by weighing the local $\Omega_P(r)$ with the local density enhancement $\left|c_m\right|$, so that the densest parts of the spiral contribute the most to the pattern speed. The corotation radius of the models with a groove is found to lie within the gap, confirming its nature as a groove mode and in accordance to what found by \cite{SellwoodKahn91}.

Since the presence of the groove diminishes the $q_D$ parameter for a given density normalization, we also set up a series of base disks (Models 12-17) with a density normalization chosen so that its mass would equal that of Models 2-7. For a given $q_D$, in the presence of the groove the $m = 2$ mode exponentiates about 3-4 times faster than the base disk. Models 14-17, despite having equal disk mass and similar $Q_{min}$ than their groovy counterparts, do not show an exponential phase during the duration of the simulation. A growing groove mode is detectable down to $q_D = 0.08$ (Model 6), but does not show a resolvable exponential phase for $q_D = 0.06$ (Model 7). The $m=2$ groove instability in Model 6 is growing about as fast as Model 13, which is four times as massive. 

\begin{figure}
\plotone{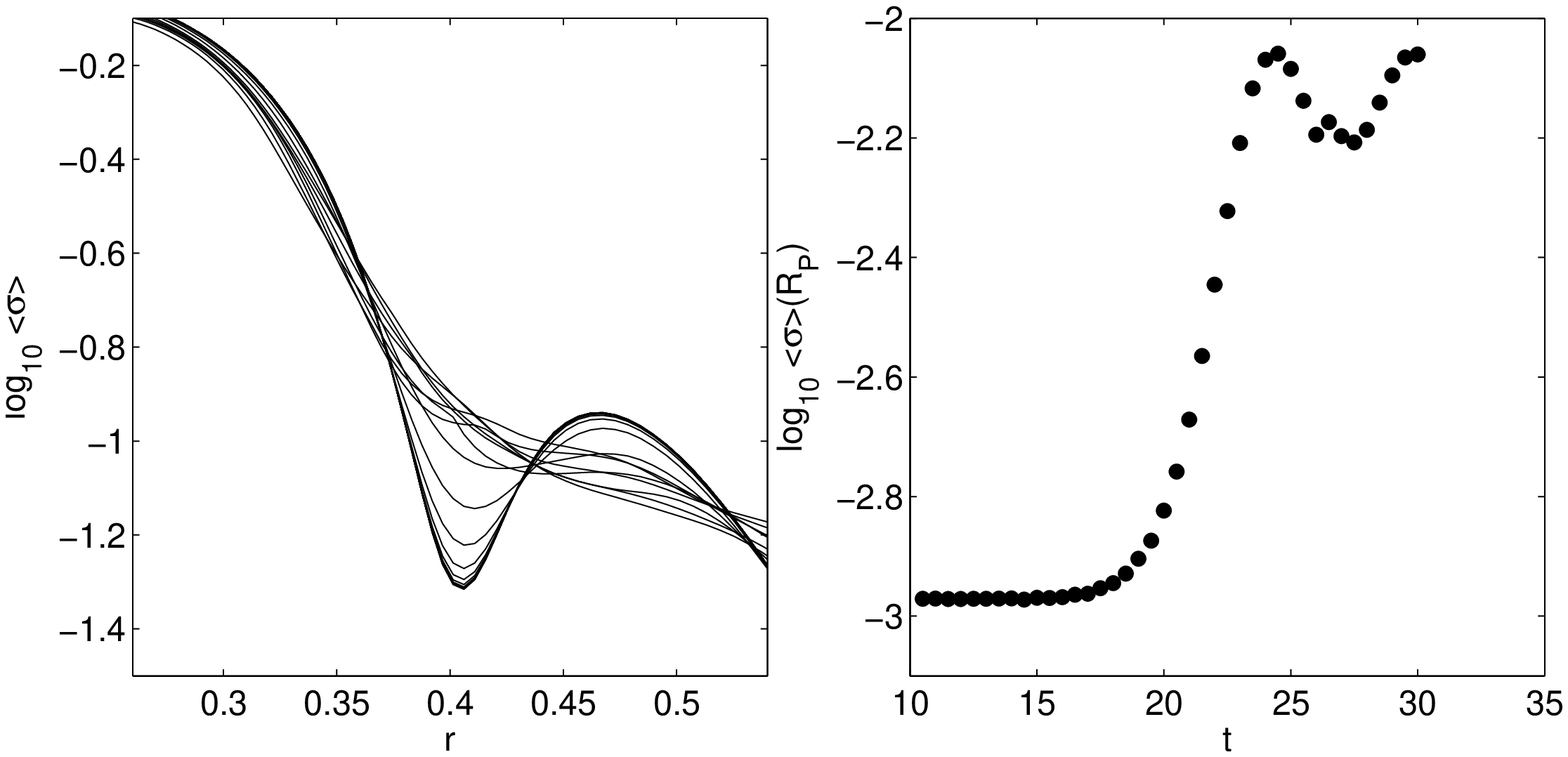}\\
\plotone{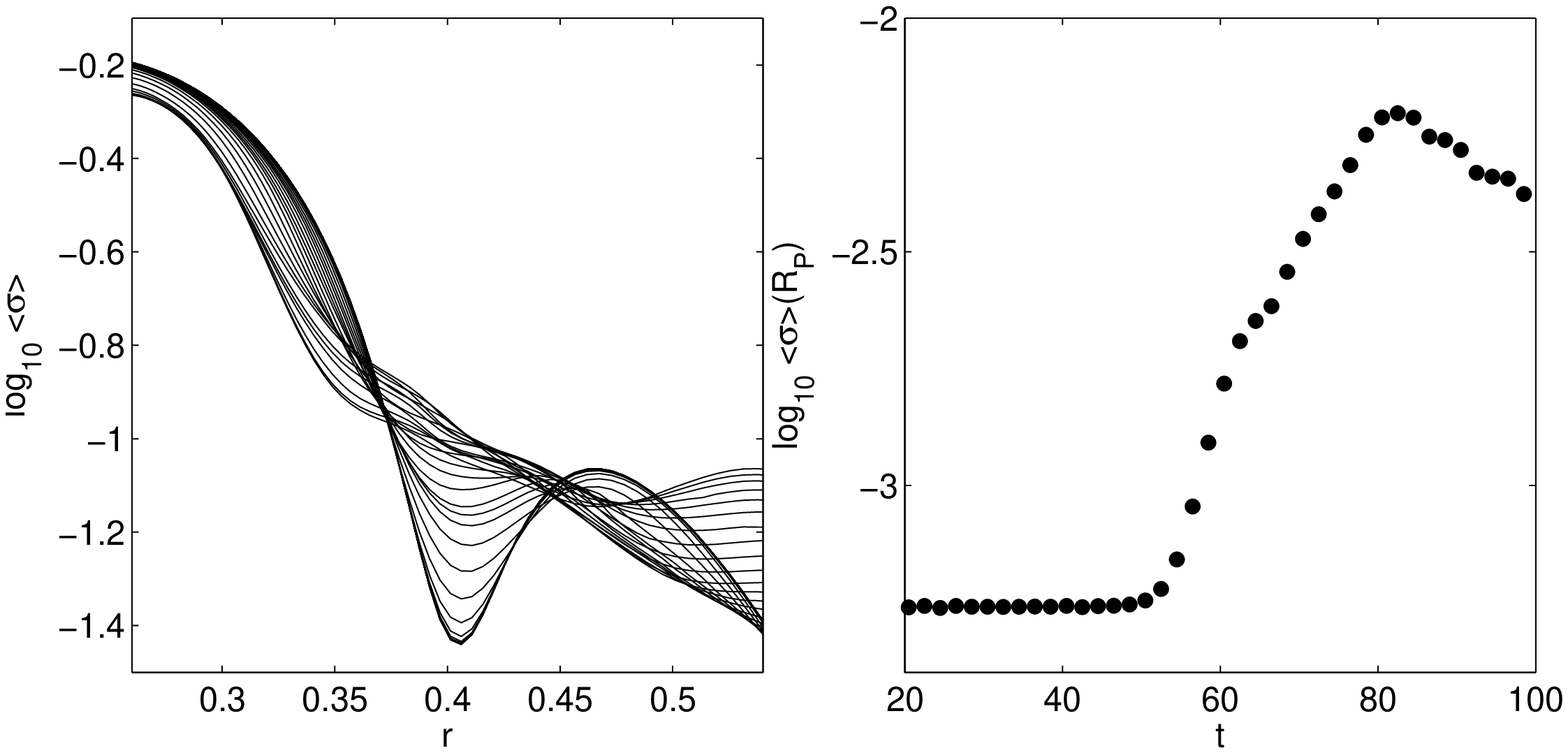}
\label{fig:alpha}
\caption{Evolution of azimuthally averaged density profile and density at the gap for models 5 (top panel) and 6 (bottom panel).}
\end{figure}

The natural outcome for the unstable models is the transport of mass and angular momentum, resulting in the filling of the gap. The evolution of the azimuthally averaged density profile for the two lowest-mass unstable models (Model 5 and 6) is shown in Figure \refp{fig:alpha}. An approximate estimate for the effective $\alpha$-type viscosity coefficient \citep{Shakura1973} is derived by a procedure similar to that employed in LR96, although the effective viscosity given by GIs is not well characterized by a local prescription. We solved the diffusion-type equation for the time-dependent surface density and compared it with the evolution in the hydrodynamical simulation, once the spiral pattern has established itself. This simple estimation yields $\alpha \sim 0.16$ for $q_D = 0.13$ and $\alpha \sim 0.04$ for $q_D = 0.08$, and agrees approximately with the timescale for closing the gap via viscous diffusion.

\section{Discussion and conclusion}
In this letter, we have found that disk self-gravity may play a significant and as-yet largely unstudied role in disks in which a planet has opened a gap in the surface density profile of the disk. Gravitational torques from massive protoplanets necessarily impart a surface density gap in a disk, and the formation of a gap provides the structure needed both for a feedback amplifier \citep[e.g.][]{Toomre81} as well as for the groove mode (SL89).  In essence, a gap provides a pressure gradient which can locally reduce the effectiveness of the epicyclic frequency in stabilizing the disk against its own self gravity, thus allowing instability at low surface densities. In the absence of a perturbing planet, the nonlinear outcome of a groove instability would be to destroy the sharp density gradient that promoted the instability in the first place. In a planet-forming disk, however, the embedded Jovian planet will exert torques whose net effect is to maintain the gap. The competition between the GI-induced  torques and the planetary torques may thus lead to a significant modification of the criterion for gap opening (as evidenced by the large effective viscosity found in our experiments), which in turn can have a significant effect on the resulting migration and growth of the protoplanet. We have found that this effect is likely to be relevant even when the disk mass is lower than the $q_D > 0.1 - 0.2$ value at which significant GIs are generally thought to occur \citep[see, e.g.][LR96]{ShuTremaine90, Boss97}. By comparison, the typical mass within 30 AU assumed for the Minimum Mass Solar Nebula yields $q_D \approx 0.01$ (this likely underestimates by a factor of 3). Further afield, \citet{Andrews2005} surveyed 153 young stellar objects in the Taurus Aurigae star-forming complex, and found a median disk mass of $q_D = 0.5\%$. While these disks span a variety of ages, and show a variety of masses, it is not clear how massive the average protoplanetary disks is at the time when cores enter the rapid gas accretion phase, and whether the mass might be enough to trigger the instability studied here. It must also be kept in mind that while the idealized disks in the simulations shown here display no GI for $q_D < 0.08$, the limited numerical resolution and high intrinsic numerical viscosity allows us to identify only modes with intrinsically rapid growth rates. Further work with realistic disk models is required to find the true minimum mass for the instability. 
We are eager to continue this analysis by (1) including the planetary potential in both our linear analysis and in our nonlinear simulations, (2) carrying out the simulations at higher numerical resolution, and (3) adopting more realistic disk models.
\acknowledgements
We thank Drs. F. Adams and D.N.C. Lin for useful discussions, and our anonymous referee for the encouraging and thorough critique. This research was supported
by the NSF through CAREER Grant AST-0449986, and by the NASA Origins of Solar Systems
Program through grant NNG04GN30G. Additional movies and color plots are available at \url{http://www.ucolick.org/\~{}smeschia/disks}.

\clearpage
\onecolumn

\begin{deluxetable}{lllllll}
\tablehead{\colhead{Model}& \colhead{$q_D$} & \colhead{$A$} & \colhead{$Q_{min}$} & \colhead{$m$} & \colhead{$\gamma_{lin}$ ($\bar{\gamma}_{nl}$)} & \colhead{$\Omega_P$} ({$\bar{\Omega}_P$})}
\tablecaption{The table lists growth rate ($\gamma_{lin}$, $\bar{\gamma}_{nl}$) and pattern speed ($\Omega_P$, $\bar{\Omega}_P$) as measured respectively by the linear code and the full hydrodynamical simulation for the various disk masses ($q_D = M_D / M_{star}$) and groove depths ($A$) considered.}
\startdata
1		&1		&	0	&	1.21	&	2	&	1.21	 (0.90)		& 2.78 (2.69)\\
2		&0.63	&	0.90	&	0.67	&	2	&	2.31	 (2.14)		&  3.41 (3.37)\\
3		&0.32	&	0.90&	1.31&	2	&	1.43	 (1.16)		& 3.18 (3.05)\\	
4		&0.16	&	0.90	&	1.92&	2	&	0.73	 (0.65)		& 3.04 (3.00)\\	
5		&0.13	&	0.90	&	2.03	&	2	&	0.56	 (0.50) &	3.02 (2.89) \\
6	 	&0.08	&	0.90	&	2.24	&	2	&	($0.28^{a,b}$) 	& (2.01) \tablenotemark{\dagger}\\
7		&0.06	&	0.90	&	2.76	&	2	&	\tablenotemark{\ddagger}	  \\
12		&0.63	&	0	&	1.30	&	2	&	0.84 (0.70)		& 2.27 (2.22)\\
13		&0.32	&	0	&	1.52	&	2	&	0.37 (0.16, 0.31)	 & 2.03 (2.22)\\	
14		&0.16	&	0	&	1.86	&	2	&	\tablenotemark{\ddagger}\\	
15		&0.13	&	0	&	2.08	&	2	&	\tablenotemark{\ddagger}\\
16		&0.08	&	0	&	2.26&	2	&	\tablenotemark{\ddagger}\\
17		&0.06	&	0	&	2.65	&	2	&	\tablenotemark{\ddagger}
\enddata
\label{tab:models}
\tablenotetext{\dagger}{This mode has two linear phases (\emph{a}, \emph{b}) and two ``saturation'' phases. The growth rates and pattern speed reported are measured from the hydro simulation.}
\tablenotetext{\ddagger}{These models do not show appreciable mode growth in either the linear or the fluid simulation.}
\end{deluxetable}


\end{document}